\documentclass[reprint,onecolumn,aip,jmp,nofootinbib,11pt]{revtex4-1}
\usepackage{amsmath}
\usepackage{amsfonts}
\usepackage{amssymb}
\usepackage{graphicx}
\usepackage{dcolumn}
\usepackage{bm}
\usepackage[colorlinks = true, allcolors = blue]{hyperref}
\usepackage{float}
\begin{document}
\title{Conditional electron confinement in graphene via smooth magnetic fields}
\author{Dai-Nam Le}
 \email{ledainam@tdt.edu.vn}
\affiliation{Atomic~Molecular~and~Optical~Physics~Research~Group,~Ton~Duc~Thang~University,~Ho~Chi~Minh~City,~Vietnam}
 \affiliation{Faculty of Applied Sciences, Ton Duc Thang University,~Ho~Chi~Minh~City,~Vietnam} 
\author{Van-Hoang Le}
\email{hoanglv@hcmup.edu.vn}
 \affiliation{Department of Physics, Ho Chi Minh City University of Pedagogy,\\ 280~An Duong Vuong St.,~Dist. 5,~Ho Chi Minh City,~Vietnam}
\author{Pinaki Roy}
\email{pinaki@isical.ac.in}
\affiliation{Physics and Applied Mathematics Unit, Indian Statistical Institute, Kolkata-700108, India}
\begin{abstract}
\begin{center}
{\bf Abstract}
\end{center}
In this article we discuss confinement of electrons in graphene via smooth magnetic fields which are finite everywhere on the plane. We shall consider two types of magnetic fields leading to systems which are conditionally exactly solvable and quasi exactly solvable. The bound state energies and wavefunctions in both cases have been found exactly.
\end{abstract}
\date{\today}

\maketitle
\section{Introduction}
In recent years graphene which is a sheet of carbon atom in honeycomb lattice \cite{g1,g2,g3} has drawn widespread attention because of its possible applications in various devices. The dynamics of charge carriers or electrons in graphene is described by the $(2+1)$ dimensional massless Dirac equation, except that the electrons move with the much smaller Fermi velocity $v_F=10^6~m/s$ instead of the velocity of light $c$. For graphene to have practical applications one of the most important problem is controlling or confining the electrons. Attempts have been made to confine electrons e.g, by using position dependent mass \cite{peres}, modulating Fermi velocity \cite{pr,down}, electrostatic fields or magnetic fields. However, confinement using electrostatic fields is usually difficult although zero energy states \cite{zero1,zero2,zero3,zero4,zero5}and sometimes some states of non zero energy \cite{qesgraphene} can be found using different field configurations. On the other hand magnetic confinement of electrons has been studied by many authors. For example, square well magnetic barrier \cite{martino1,martino2}, radial magnetic field \cite{maksym}, decaying gaussian magnetic field \cite{tkg}, hyperbolic magnetic fields \cite{murguia}, inhomogeneous magnetic fields \cite{masir,pratim,downing1,downing2,esh}, one dimensional magnetic fields leading to solvable systems \cite{kuru}, etc. have been used to create bound states in graphene. In particular, of the different types of magnetic fields mentioned above, there are some smooth inhomogeneous magnetic fields \cite{pratim,downing1,downing2} for which the pseudo spinor components satisfy equations with quasi exactly solvable effective potentials \cite{tur}. In this context, it may be noted that inhomogeneous magnetic field profiles can be produced in many ways e.g, using ferromagnetic materials \cite{inhomo1}, non planar substrate \cite{inhomo2}, integrating superconducting elements \cite{inhomo3} etc. In the present paper, our objective is to search for smooth everywhere finite magnetic fields which produce conditionally exactly solvable effective potentials \cite{dutra,junker} i.e, potentials which admit exact solutions when parameter(s) of the model assume particular values. More precisely, it will be shown that the electrons remain confined for certain values of the magnetic quantum number while for other values of hte magnetic quantum number they enter the deconfining phase. We shall also explore the possibility of obtaining quasi exactly solvable systems when some of the constraints on the parameters are relaxed. The organization of the paper is as follows: in section \ref{forma} we shall present the formalism; in section \ref{CES} we shall obtain several magnetic fields which leads to conditionally exactly solvable systems; in section \ref{QES} we shall examine under what conditions the magnetic fields produce quasi exactly solvable systems and finally section \ref{con} is devoted to a conclusion.
 \section{Formalism}\label{forma}
The dynamics of quasi particles in graphene is governed by the Hamiltonian
\begin{eqnarray}
\label{eqn:Hamilton}
\hat{H} = v_F \vec{\sigma} \cdot \vec{\hat{\pi}} && = v_F \vec{\sigma} \cdot \left( \vec{\hat{p}} + \vec{A} \right) = v_F \left( 
\begin{matrix}
0 & \hat{\Pi} _{-} \\
\hat{\Pi} _+ & 0
\end{matrix} \right) ,
\end{eqnarray}
where $v_F$ is the Fermi velocity, $\sigma = ( \sigma _x , \sigma _y )$ are Pauli matrices, and
\begin{equation}
\hat{\Pi}_{\pm} = \hat{\pi} _x \pm i \hat{\pi} _y = (\hat{p}_x + A_x) \pm i (\hat{p}_y + A_y).
\end{equation}
We now choose the vector potentials to be of the form
\begin{equation}
A_x=yf(r),~~~~A_y=-xf(r)
\end{equation}
where the specific form of the function $f(r)$ will be chosen later. With the above choice of the vector potentials, the magnetic field is given by
\begin{equation}
B_z=-2f(r)-rf'(r).
\end{equation} 
The eigenvalue equation
\[ \hat{H} \psi = E \psi , \]
where $\psi=(\psi_1,\psi_2)^T$ is a two component pseudospinor, can be written as
\begin{eqnarray}
\hat{\Pi} _{-} \psi_2 &&= \epsilon \psi_1 ,\label{eqn:psi1-0}\\
\hat{\Pi} _{+} \psi_1 &&= \epsilon \psi_2 , \label{eqn:psi2-0}
\end{eqnarray} 
where $\epsilon=E/v_F$. Now eliminating $\psi_1$ in favor of $\psi_2$ (and vice-versa), the equations for the components can be written as
\begin{eqnarray}
\hat{\Pi} _{-} \hat{\Pi} _{+} \psi_1 &&= \epsilon^2 \psi_1 , \label{eqn:psi1}\\
\hat{\Pi} _{+} \hat{\Pi} _{-} \psi_2 &&= \epsilon^2 \psi_2 \label{eqn:psi2}. 
\end{eqnarray} 
Since the magnetic field is a radial one, the pseudospinor components can be taken as
\begin{equation}
\psi_1=e^{im\theta}r^{-1/2}\phi_1(r),~~~~\psi_2=e^{i(m+1)\theta}r^{-1/2}\phi_2(r),~~m=0,\pm 1,\pm 2,\cdots,
\end{equation}
where $m$ is the magnetic quantum number. Then eigenvalue equations for the components can be written as
\begin{equation}\label{eqn:psi1-a}
\displaystyle\left[-\frac{d^2}{dr^2}+\frac{m^2-\frac{1}{4}}{r^2}+ r^2 f^2 - 2(m+1)f - rf'\right] \phi_1 = \epsilon^2 \phi_1,\\
\end{equation}
\begin{equation}\label{eqn:psi2-a}
\displaystyle\left[-\frac{d^2}{dr^2} + \frac{(m+1)^2-\frac{1}{4}}{r^2}+ r^2 f^2 - 2mf + rf'\right]   \phi _2 = \epsilon^2 \phi _2. 
\end{equation}

Before closing this section, we note that the intertwining relations \eqref{eqn:psi1-0} and \eqref{eqn:psi2-0} can also be written in terms of polar coordinates and are given by
\begin{equation}\label{intert}
\begin{array}{l}
\displaystyle\left( \dfrac{\partial}{\partial r}- \dfrac{m+\frac{1}{2}}{r} + r f \right) \phi _1 = i \epsilon \phi _2,\\
\displaystyle\left( \dfrac{\partial}{\partial r}+ \dfrac{m+\frac{1}{2}}{r} - r f \right) \phi _2 = i \epsilon \phi _1.
\end{array}
\end{equation}
The set of intertwining relations \eqref{intert} is particularly important since knowing solution of one of the two equations \eqref{eqn:psi1-a} or \eqref{eqn:psi2-a}, the other can be obtained through the above relations.

\section{Conditionally exactly solvable magnetic fields}\label{CES} 
Here we shall consider several conditionally exactly solvable magnetic field profiles i.e, magnetic fields for which all or some bound state solutions can be found {\it only} when the parameters of the model assume some specific values. To this end we choose the function $f(r)$ to be of the form
\begin{equation}
\label{eqn:consider-f}
f(r) = \dfrac{\lambda}{2} + \sum _{i=1}^{N} \dfrac{2 g_i}{1 + g_i r^2},~~\lambda>0,~~g_1,g_2,...,g_N>0. 
\end{equation}
Then the resulting magnetic field is given by
\begin{equation}
\label{eqn:consider-B}
B_z (r) = - \lambda - \sum _{i=1}^{N} \dfrac{4 g_i}{(1 + g_i r^2)^2}.
\end{equation}

From equation \eqref{eqn:consider-B} it can be observed that the magnetic field is everywhere finite with a maximum value of $-\lambda$ and a minimum of $-\lambda-4\sum_{i=0}^Ng_i$. We  shall now consider different values of $N$ and examine if the corresponding magnetic field can support bound states when the parameters assume some particular values. 
\subsection{$N=1$}
In this case the magnetic field becomes
\begin{equation}
\label{eqn:B-1}
B_z (r) = - \lambda - \dfrac{4 g_1}{(1 + g_1 r^2)^2},
\end{equation}
and the profile of this field can be seen in Fig 1.
\begin{figure}[H]
\begin{center}
\includegraphics[width=0.4 \textwidth]{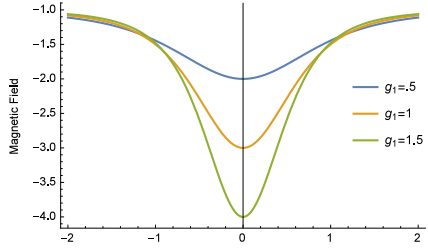}
\caption{\label{fig:magN1}Magnetic field profile for $N=1, \lambda=1$.}
\end{center}
\end{figure}
Then, from \eqref{eqn:psi1-a} and \eqref{eqn:psi2-a} the equations for the components $\phi _1$ and $\phi _2$ can be obtained as
\begin{equation}\label{eqn:psi1-1}
\displaystyle \left[-\dfrac{d^2}{dr^2} +\displaystyle\dfrac{m^2-\frac{1}{4}}{r^2} + \frac{\lambda ^2 r^2}{4} - \frac{2Z - 4g_1}{1 + g_1 r^2} - \frac{8 g_1}{(1 + g_1 r^2 )^2} \right] \phi_1 = \left[ \epsilon^2 + \lambda (m-1)\right] \phi_1,
\end{equation} 
\begin{equation}\label{eqn:psi2-1}
\displaystyle\left[- \dfrac{d^2}{dr^2} + \dfrac{\left(m+1\right)^2-\frac{1}{4}}{r^2}+ \frac{\lambda ^2 r^2}{4}  - \frac{2Z}{1 + g_1 r^2}\right]\phi _2 \\
= \left[ \epsilon^2 + \lambda (m-2)\right] \phi _2, 
\end{equation}
where $Z = 2 m g_1 + \lambda$.

\begin{figure}[H]
\begin{center}
\includegraphics[width = 0.6 \textwidth]{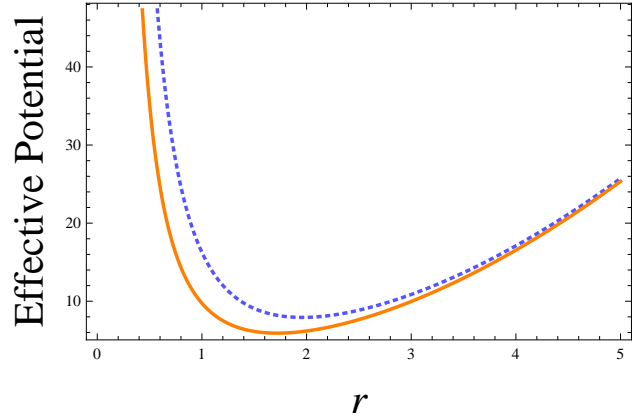}
\caption{\label{fig:2} Plots of effective potentials in Eq.\eqref{eqn:psi1-1} (dotted curve) and Eq.\eqref{eqn:psi2-1} (solid curve) for $\lambda=2, M=4$ and $Z=0$.}
\end{center}
\end{figure}

\begin{figure}[H]
\begin{center}
\includegraphics[width = 0.6 \textwidth]{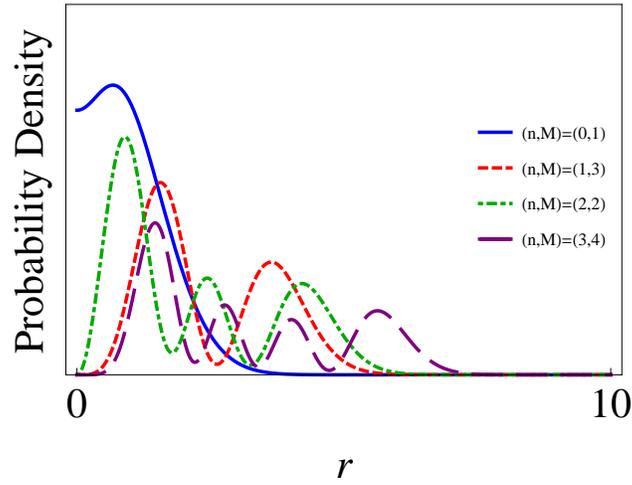}
\caption{\label{fig:3} Plots of the probability density for some values of $n$ and $M$.}
\end{center}
\end{figure}

\paragraph*{Conditional exact solutions:}
Let us now consider equation \eqref{eqn:psi2-1} for the lower component. This equation can be interpreted as the radial Schr\"odinger equation for a particle moving in a two dimensional nonpolynomial oscillator potential. Next, we choose the parameter $g_1$ in such a way that the nonpolynomial part vanishes i.e, \footnote{Note that if $\lambda<0$, solutions can be obtained in a similar way for the sector $m>0$}
\begin{equation}
\label{eqn:N1-cond}
Z=0\Rightarrow g_1 =- \lambda / 2 m.
\end{equation}
Now recalling that $g_1$ and $\lambda$ are always positive, the admissible values of $m$ are $m<0$. With $g_1$ as given above, equation \eqref{eqn:psi2-1} becomes the radial Schr\"odinger equation for the two-dimension isotropic harmonic oscillator :
\begin{equation}
\label{eqn:psi2-1-harmo}
\left[ - \dfrac{d^2}{d r^2} + \dfrac{(M-1)^2 - \frac{1}{4}}{r^2} + \dfrac{\lambda^2 r^2}{4} \right] \phi_2 = \left[ \epsilon^2 - \lambda (M + 2) \right] \phi _2,~~M=-m.
\end{equation} 
It may be pointed out the effective potential becomes that of the radial harmonic oscillator only when $g_1$ assumes the particular value given by (\ref{eqn:N1-cond}). The eigenvalues and the corresponding wave functions of \eqref{eqn:psi2-1-harmo} are standard and are given by :
\begin{equation}
\label{eqn:E-1-harmo}
E_{n,M} = \pm v_F\sqrt{2 \lambda (n + M + 1)},~n=0,1,2,\cdots,~M=1,2,\cdots
\end{equation}
\begin{equation}
\label{eqn:psi2-1-harmo-sol}
\phi _2 (r) \sim r^{M-1} e^{-\lambda r^2 / 4} \mathcal{L} _{n}^{M-1} \left( \lambda r^2 / 2 \right) ,
\end{equation}
where $\mathcal{L} _n^M (x)$ is the associated Laguerre polynomial. Then, the lower component of the pseudospinor wave function $\psi _2$ is
\begin{equation}
\label{eqn:psi2-1-harmo-lower}
\psi _2 (r, \theta ) \sim r^{M-1} e^{-\lambda r^2 / 4} \mathcal{L} _{n}^{M-1} \left( \lambda r^2 / 2 \right) e^{i (1 - M) \theta } .
\end{equation}
From the intertwining relation \eqref{intert}, the upper component can be determined through the lower component \eqref{eqn:psi2-1-harmo-lower} and we obtain the pseudospinor wave function:
\begin{eqnarray}
\label{eqn:psi-1-harmo-spinor}
\psi (r, \theta ) \sim \; &&
r^{M-1} e^{-\lambda r^2 / 4} e^{- i M \theta} \times \nonumber\\
&& \times \left(
\begin{matrix}
i \epsilon^{-1}_{n,M} \lambda r \left[ \mathcal{L}_n^{M}(\lambda r^2 /2 ) + \dfrac{2}{\lambda r^2 + 2 M} \mathcal{L}_n^{M-1}(\lambda r^2 /2 ) \right]
\\
e^{i \theta } \mathcal{L} _{n}^{M-1} \left( \lambda r^2 / 2 \right)
\end{matrix} 
\right).
\end{eqnarray}  
It may be noted that as the magnetic quantum number decreases and becomes $M<1$, the electrons enter the deconfining phase and are no longer confined. In Figs \ref{fig:2} and \ref{fig:3} we have presented plots of the effective potentials in Eqs.\eqref{eqn:psi1-1} and \eqref{eqn:psi2-1} for $Z=0$ and probability density for different values of the parameters.
\subsection{$N=2$}
\begin{figure}[H]
\begin{center}
\includegraphics[width=0.4 \textwidth]{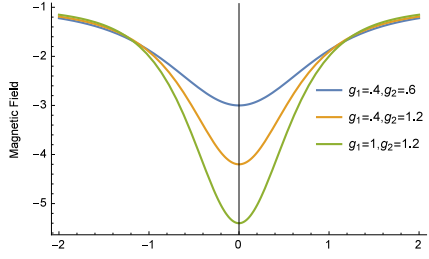}
\caption{\label{fig:magN2}Magnetic field profile for $N=2, \lambda=1$.}
\end{center}
\end{figure}

Here we shall consider a more general magnetic field and put $N = 2$ in \eqref{eqn:consider-f} and obtain:
\begin{equation}\label{eqn:f-2}
f(r) = \dfrac{\lambda}{2} + \dfrac{2 g_1}{1 + g_1 r^2} + \dfrac{2 g_2}{1 + g_2 r^2},
\end{equation}
and the corresponding magnetic field is given by
\begin{equation}\label{eqn:B-2}
B_z (r) = - \lambda - \dfrac{4 g_1}{(1 + g_1 r^2)^2} - \dfrac{4 g_2}{(1 + g_2 r^2)^2}.
\end{equation}
Fig \ref{fig:magN2}. shows the profile of the magnetic field for different values of the parameters.

From \eqref{eqn:psi1-a} and \eqref{eqn:psi2-a}, the equations for $\phi _1$ and $\phi _2$ can be obtained as
\begin{equation}\label{eqn:psi1-2}
\begin{array}{l}
\left[-\dfrac{d^2}{dr^2} +\dfrac{m^2-1/4}{r^2} + \dfrac{\lambda ^2 r^2}{4} - \dfrac{2Z_1 - 4g_1}{1 + g_1 r^2}
 - \dfrac{8 g_1}{(1 + g_1 r^2 )^2}- \dfrac{2Z_2 - 4g_2}{1 + g_2 r^2} - \dfrac{8 g_2}{(1 + g_2 r^2 )^2} \right] \phi_1 \\
= \left[ \epsilon^2 + \lambda (m-3)\right] \phi_1,
\end{array}
\end{equation}
\begin{equation}\label{eqn:psi2-2}
\left[- \dfrac{d^2}{dr^2} +\dfrac{\left(m+1\right)^2-1/4}{r^2}+ \dfrac{\lambda ^2 r^2}{4}  - \dfrac{2Z_1}{1 + g_1 r^2} - \dfrac{2Z_2}{1 + g_2 r^2},  \right]   \phi _2 = \left[ \epsilon^2 + \lambda (m-4)\right] \phi _2, 
\end{equation}
where $Z_1 = 2 m g_1 + \lambda + 4 g_1 g_2 /(g_1 - g_2)$ and $Z_2 = 2 m g_2 + \lambda + 4 g_2 g_1 /(g_2 - g_1)$.

\begin{figure}[H]
\begin{center}
\includegraphics[width = 0.6 \textwidth]{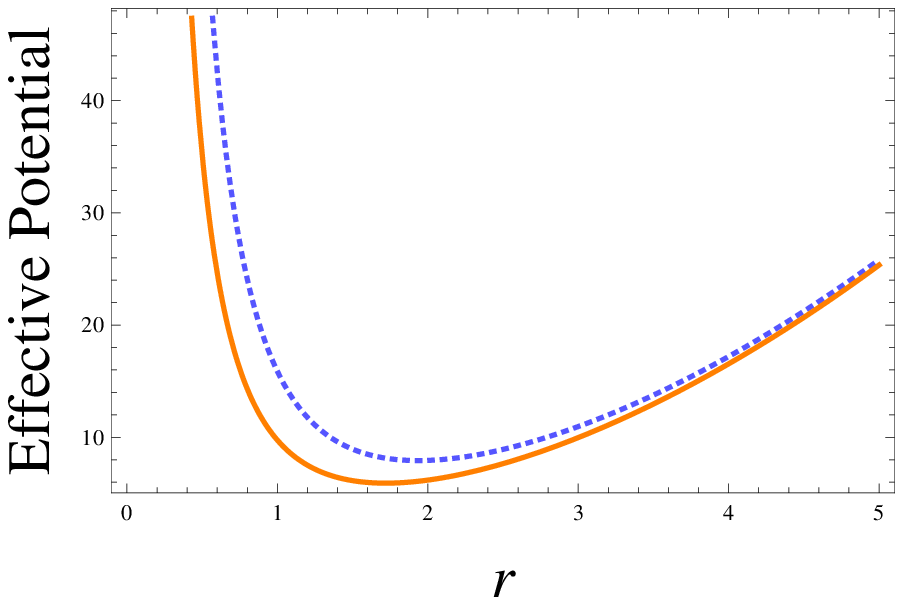}
\caption{\label{fig:5}Plots of effective potentials in Eq.\eqref{eqn:psi1-2} (dotted curve) and Eq.\eqref{eqn:psi2-2} (solid curve) for $\lambda=2, M=4$ and $Z_1=0=Z_2$.}
\end{center}
\end{figure}

\begin{figure}[H]
\begin{center}
\includegraphics[width = 0.6 \textwidth]{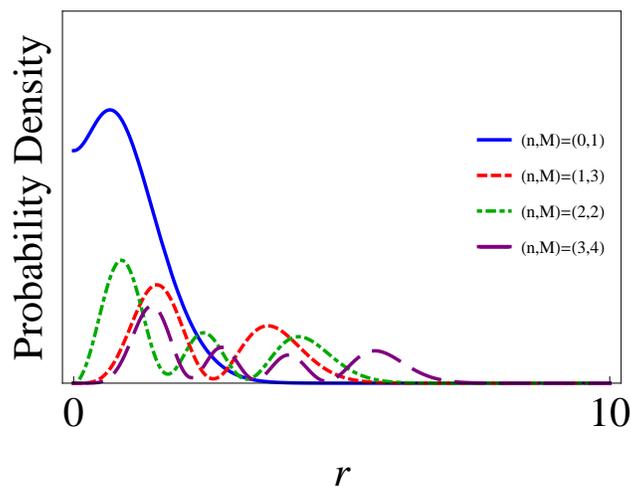}
\caption{\label{fig:6}Plots of the probability density for some values of $n$ and $M$.}
\end{center}
\end{figure}

\paragraph*{Conditional exact solutions:} As in the previous example, we consider equation \eqref{eqn:psi2-2} for the lower component. In this case, the nonpolynomial interaction is a more general one and consists of two terms representing the nonlinearities. However, the potential reduces to the two dimensional harmonic oscillator potential if $Z_1=0=Z_2$ i.e,
\begin{equation}
\begin{array}{l}
2 m g_1 + \lambda + 4 g_1 g_2 /(g_1 - g_2)=0,\\
2 m g_2 + \lambda - 4 g_1 g_2 /(g_1 - g_2)=0.
\end{array}
\end{equation}
The solution of the above set of coupled equation is given by
\begin{equation}
\begin{array}{l}
 g_1 = \dfrac{\lambda}{2 (M+1 + \sqrt{M+1})}>0,\\ 
 g_2 = \dfrac{\lambda}{2 (M+1 - \sqrt{M+1})}>0,
 \end{array}
 \end{equation}
where $M = 1, 2, 3, \cdots$ and is related to the the magnetic quantum number by $m = - M$. With the above choice of $g_{1,2}$, equation \eqref{eqn:psi2-2} immediately becomes radial Schr\"odinger equation of an isotropic harmonic oscillator :
\begin{equation}
\label{eqn:psi2-2-harmo}
\left[ - \dfrac{d^2}{d r^2} + \dfrac{(M-1)^2 - 1/4}{r^2} + \dfrac{\lambda^2 r^2}{4} \right] \phi_2 = \left[ \epsilon^2 - \lambda (M + 4) \right] \phi _2 .
\end{equation} 
The energy and corresponding eigenfunctions are given by
\begin{equation}
\label{eqn:E-2-harmo}
E_{n,M} = \pm v_F\sqrt{2 \lambda (n + M + 2)},~~n=0,1,2\cdots,~~M=1,2,3,\cdots
\end{equation}
\begin{equation}
\label{eqn:psi2-2-harmo-sol}
\phi _2 (r) \sim r^{M-1} e^{-\lambda r^2 / 4} \mathcal{L} _{n}^{M-1} \left( \lambda r^2 / 2 \right).
\end{equation}
Then using the intertwining relation \eqref{intert}, the pseudospinor can be obtained as
\begin{eqnarray}
\label{eqn:psi-2-harmo-spinor}
\psi_{n,M} (r, \theta ) \sim \; &&
r^{M-1} e^{-\lambda r^2 / 4} e^{- i M \theta} \times \nonumber\\
&& \times \left(
\begin{matrix}
i \epsilon^{-1}_{n,M} r \left[ \lambda \mathcal{L}_n^{M}(\lambda r^2 /2 ) + \left( \dfrac{2 g_1}{1 + g_1 r^2} + \dfrac{2 g_2}{1 + g_2 r^2} \right) \mathcal{L}_n^{M-1}(\lambda r^2 /2 ) \right]
\\
e^{i \theta } \mathcal{L} _{n}^{M-1} \left( \lambda r^2 / 2 \right)
\end{matrix} 
\right) .
\end{eqnarray} 
As before the electrons remain confined  for $M\geq 1$. In Figs \ref{fig:5} and \ref{fig:6} we have plotted the effective potentials in Eq.\eqref{eqn:psi1-2} and Eq.\eqref{eqn:psi2-2} for $Z_1=0=Z_2$ and probability density for several values of the parameters.


Before we conclude this section, let us examine degeneracy of the eigenvalues. In both the cases considered above it is seen that the ground state $E_{0,1}$ is non degenerate while all other states are degenerate. For example, $E_{1,1}=E_{0,2},~E_{1,2}=E_{2,1}=E_{0,3},~E_{1,3}=E_{3,1}=E_{2,2}=E_{0,4}$ and so on. Thus degeneracy of the level $E_{n,M}$ is $(n+M)$.

\section{Quasi exact solutions}\label{QES}
Here we shall explore whether or not the magnetic fields considered in the previous section may produce effective potentials which are quasi exactly solvable, that is, potentials for which only some solutions can be found analytically. It will be seen that one may indeed find some exact solutions even when some of the constraints are relaxed.

Before considering the quasi exact solutions for $N =1$ and $N=2$, we note that from here now the notation
\[ M = \left| m + \dfrac{1}{2} \right| - \dfrac{1}{2} = \left\{
\begin{array}{ll}
m & \text{if } m \geq 0,\\
-(m+1) & \text{if } m < 0,
\end{array}
\right. \]
is used to separate the cases of $m \geq 0$ and of $m <0$.

\subsection{$N=1$.}
Let us first consider the case $N=1$.       
In this case the pseudospinor components satisfy equations \eqref{eqn:psi1-1} and \eqref{eqn:psi2-1} and we obtained conditionally exact solutions under the condition $Z=0$. The question which comes up immediately is the following: Can we still find bound states when $Z\neq 0$? Before answering this question, we would like to note that the parameter $g_1$ should always be positive since for $g_1<0$, the magnetic field becomes singular and the potentials
\begin{equation}\label{pot}
\begin{array}{l}
V_1(r)=\dfrac{\lambda ^2 r^2}{4} - \dfrac{2Z - 4g_1}{1 + g_1 r^2} - \dfrac{8 g_1}{(1 + g_1 r^2 )^2},\\
V_2(r)=\dfrac{\lambda ^2 r^2}{4}  - \dfrac{2Z}{1 + g_1 r^2}
\end{array}
\end{equation}
also become singular and they may not share the same spectrum. In order to obtain bound states, let us first consider the zero energy ones. For $Z\neq 0$ and $g_1>0$ the zero energy solutions can be easily obtained from equation \eqref{intert} for $\lambda < 0$ and they are given by
\begin{equation}
\label{eqn:psi-1-quasi-spinor}
E=0,~~~~\psi_{0,M} (r , \theta) \sim e^{-iM \theta }\left( 
\begin{matrix}
0\\
r^{M} (1+ g_1 r^2) e^{ \lambda r^2 / 4} 
\end{matrix}
\right),~~M=0,1,2,\cdots .
\end{equation}
On the other hand for $\lambda>0$, the zero energy solutions are given by 
\begin{equation}\label{eqn:psi-1-quasi-spinor-b}
E=0,~~~~\psi_{0,M}(r,\theta)\sim e^{iM\theta}\left(\begin{array}{c}r^{M}e^{-\lambda r^2/4}/(1+g_1r^2)\\0\end{array}\right),~~M=0,1,2,\cdots.
\end{equation}
Note that in both the cases the solutions $(1)$ are infinitely degenerate with respect to the quantum number $M$, $(2)$ exist only for some values of the magnetic quantum number.

To determine non zero energy solutions, we choose $V_2(r)$ and consider a wave function of the form 
\begin{equation}\label{phi2N=1}
\phi _2 \sim r^{|m+1|+1/2} (1 + g_1 r^2 ) e^{- |\lambda | r^2 / 4} \sum _{n=0}^{\mathcal{N}} c_n r^{2n}.
\end{equation}
Now substituting \eqref{phi2N=1} in \eqref{eqn:psi2-1} we find after some calculations that \cite{nam}
\begin{equation}\label{quasin=1}
E_{{\mathcal{N}},M}= \pm v_F \sqrt{|\lambda| (2 \mathcal{N} + |m+1| + (m+1)) - (|\lambda| + \lambda) (m-2)} ,
\end{equation}
\begin{equation}
\phi _2 \sim r^{|m+1|+1/2} (1 + g_2 r^2 ) e^{- |\lambda | r^2 / 4} \sum _{n=0}^{\mathcal{N}} \dfrac{(-1)^n (|m+1|)!}{2^n n! (n+|m+1|)!} D_n(Z,E) r^{2n},
\end{equation}
where
\begin{equation}
\label{eqn:Dn-quasi-1}
\begin{array}{l}
D_n (Z, E) =  \left|
\begin{matrix}
a_1  & b_1    & 0      & \cdots & 0      \\       
c_1    & a_2  & b_2    & \ddots & \vdots \\
0      & c_2    & a_3  & \ddots &  0     \\
\vdots & \ddots & \ddots & \ddots &  b_{n-1} \\
0      & \cdots & 0      & c_{n-1}     & a_n    
\end{matrix}
 \right| ,\\ 
a_k = 2 g_2 k (k+|m+1|) + Z + |\lambda | (\mathcal{N}+2 - k),~~ b_k = 2g_2 k (k+|m+1|),~~c_k = |\lambda| (\mathcal{N}+1-k).
\end{array}
\end{equation}

Clearly for the energy levels \eqref{quasin=1} to be admissible ones $D_n(Z,E)=0$ for some $n>1$. However, we have not found any such solutions consistent with the constraint $g_1>0$.

\subsection{$N=2$} In this case, the potentials in Eqs. \eqref{eqn:psi1-2} and \eqref{eqn:psi2-2} are given by
\begin{equation}\label{v2-1}
\displaystyle V_1(r)=\dfrac{\lambda ^2 r^2}{4} 
- \dfrac{2Z_1 - 4g_1}{1 + g_1 r^2} - \dfrac{8 g_1}{(1 + g_1 r^2 )^2}  - \dfrac{2Z_2 - 4g_2}{1 + g_2 r^2} - \dfrac{8 g_2}{(1 + g_2 r^2 )^2},
\end{equation}
\begin{equation}\label{v2-2}
\displaystyle V_2(r)=\dfrac{\lambda ^2 r^2}{4}  - \dfrac{2Z_1}{1 + g_1 r^2} - \dfrac{2Z_2}{1 + g_2 r^2}.
\end{equation}
A plot of these potentials are shown in Fig 7. 
\begin{figure}[H]
\begin{center}
\includegraphics[width = 0.6 \textwidth]{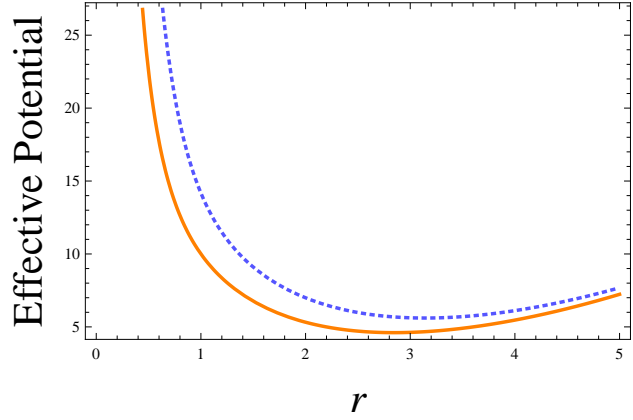}
\caption{\label{fig:7} Plots of effective potentials in Eq.\eqref{v2-1} (dotted curve) and Eq.\eqref{v2-2} (solid curve) for $\lambda=1, m=-3$ and $g_1=0.3484878472$, $g_2=0.6593973328$.}
\end{center}
\end{figure}

The zero energy states in this case can be found as in the previous example and for $\lambda<0$ are given by
\begin{equation}
\label{eqn:quasi N=2}
E=0,~~~~\psi_{0,M} (r , \theta) \sim e^{-iM \theta }\left( 
\begin{matrix}
0\\
r^{M} (1+ g_1 r^2)(1+g_2r^2) e^{\lambda r^2 / 4} 
\end{matrix}
\right),~~M=0,1,2,\cdots ,
\end{equation}
while for $\lambda>0$ are given by
\begin{equation}
E=0,~~~~\psi_{0,M}(r,\theta)\sim e^{iM \theta}\left(\begin{array}{c}r^{M}e^{-\lambda r^2/4}/(1+g_1r^2)(1+g_2r^2)\\0\end{array}\right),~~M=0,1,2,\cdots
\end{equation}
In order to find whether any of the potentials above can support quasi exact solutions, we first choose $V_2(r)$ and consider $Z_1=0$. In this case, we obtain
\begin{equation}
g_2 = \dfrac{g_1 (2m g_1 + \lambda )}{(2m - 4) g_1 + \lambda }.
\end{equation}
Then equation \eqref{eqn:psi2-2} becomes
\begin{equation}
\label{eqn:psi2-2-quasi}
\left[- \dfrac{d^2}{dr^2} + \dfrac{\left(m+1\right)^2-1/4}{r^2}+ \dfrac{\lambda ^2 r^2}{4}  - \dfrac{2Z_2}{1 + g_2 r^2} \right]   \phi _2 = \left[ \epsilon^2 + \lambda (m-4)\right] \phi _2,
\end{equation}
where $Z_2$ is given by
\begin{equation}
Z_2 = 4 m g_1 \left( 1 + \dfrac{2 g_1}{2 (m -2) g_1 + \lambda } \right) + 2 \lambda.
\end{equation}
It may be noted that equations \eqref{eqn:psi2-1} and \eqref{eqn:psi2-2-quasi} look quite similar although they can not be identified with one another since $Z\neq Z_2$. The procedure for obtaining non zero energy states is similar to the previous case and the results are
\begin{equation}
\label{eqn:E-2-quasi}
E_{{\mathcal N},M} = \pm v_F\sqrt{|\lambda | (2 \mathcal{N} + |m+1| - (m+1) + 8) + (|\lambda| - \lambda )(m-4)},
\end{equation}
\begin{equation}
\label{eqn:psi2-2-sol}
\phi _2 \sim r^{|m+1|+1/2} (1 + g_2 r^2 ) e^{- |\lambda | r^2 / 4} \sum _{n=0}^{\mathcal{N}} \dfrac{(-1)^n (|m+1|)!}{2^n n! (n+|m+1|)!} D_n r^{2n},
\end{equation}
where the expression for $D_n$ reads exactly as in \eqref{eqn:Dn-quasi-1} except that one has to make the change $Z\rightarrow Z_2$.

The bound state solutions can be obtained from the condition
\begin{equation}\label{dnN=2}
D_n(Z_2,E)=0
\end{equation}
subject to the condition $g_1,g_2>0$.
In general, this condition is a polynomial equation for $g_2$ with degree of $(\mathcal{N} +1)$.

Some specific solutions can be obtained as follows: for $\lambda > 0$ and $m = - M - 1 < 0$ ($M = 0 , 1 , 2 , \dots $), one can find exact solutions for all non-negative integers $\mathcal{N}$. For example, for $\mathcal{N} = 0$, the admissible values of $g_1$ and $g_2$ are
\[ g_1 = \dfrac{3 \lambda }{2 M + 2},  \quad g_2 = \dfrac{3 \lambda }{2 M + 5}. \]
The energy and the pseudospinor wave function now are
\begin{equation*}
E = \pm 2v_F \sqrt{2 \lambda},
\end{equation*}
\begin{equation*}
\phi _2 \sim  r^{M} e^{- \lambda  r^2 / 4} \left( 3 \lambda r^2 + 2M + 5 \right) e^{- i M \theta}  ,
\end{equation*}
and
\begin{equation*}
\psi (r , \theta ) \sim r^{M} e^{- \lambda  r^2 / 4} e^{- i M \theta} \times \left(
\begin{matrix}
i E^{-1} r \left( 3 \lambda ^2 r^2 + (2M+11) \lambda r + \dfrac{18}{3 \lambda r^2 + 2M + 2} \right) e^{- i \theta} \\
\left( 3 \lambda r^2 + 2M + 5 \right)
\end{matrix}
\right) .
\end{equation*}
In general for $\mathcal{N} > 0$, it is difficult to solve equation \eqref{dnN=2} analytically. So we have solved it numerically and a sample of the results for $\lambda = 1$ are shown in Table \ref{tab:1}.
\begin{table}[H]
\caption{\label{tab:1}Allowed values of $g_1$, $g_2$ and exact energy values $E$ for $m=-1, -2, -3, -4, -5$ and $\mathcal{N} = 1, 2, 3$.}
\scriptsize
\begin{ruledtabular}
\begin{tabular}{ccccc}
$m = - M - 1$ & $g_1$ & $g_2$ & $E$ & nodes \\
\hline
\multicolumn{5}{c}{$\mathcal{N} = 1$}\\

$-1$&$ 0.4743416490E+00$&$ 0.1790569415E+01$&$\pm 0.3162277660E+01$&$1$\\

$-2$&$ 0.3535533906E+00$&$ 0.8535533906E+00$&$\pm 0.3464101615E+01$&$1$\\

$-3$&$ 0.2834733548E+00$&$ 0.5538126093E+00$&$\pm 0.3741657387E+01$&$1$\\

$-4$&$ 0.2371708245E+00$&$ 0.4081138830E+00$&$\pm 0.4000000000E+01$&$1$\\

$-5$&$ 0.2041241452E+00$&$ 0.3224744871E+00$&$\pm 0.4242640687E+01$&$1$\\
\hline
\multicolumn{5}{c}{$\mathcal{N} = 2$}\\
$-1$&$ 0.5715576511E+00$&$ 0.2077086572E+01$&$\pm 0.3464101615E+01$&$2$\\

$-2$&$ 0.4058817282E+00$&$ 0.9555757090E+00$&$\pm 0.3741657387E+01$&$2$\\

$-3$&$ 0.3162345421E+00$&$ 0.6068766798E+00$&$\pm 0.4000000000E+01$&$2$\\

$-4$&$ 0.2596159465E+00$&$ 0.4408038959E+00$&$\pm 0.4242640687E+01$&$2$\\

$-5$&$ 0.2204606934E+00$&$ 0.3446849045E+00$&$\pm 0.4472135955E+01$&$2$\\
\hline
\multicolumn{5}{c}{$\mathcal{N} = 3$}\\
$-1$&$ 0.6674965423E+00$&$ 0.2361139093E+01$&$\pm 0.3741657387E+01$&$3$\\

$-2$&$ 0.4573940766E+00$&$ 0.1056561074E+01$&$\pm 0.4000000000E+01$&$3$\\

$-3$&$ 0.3484878472E+00$&$ 0.6593973328E+00$&$\pm 0.4242640687E+01$&$3$\\

$-4$&$ 0.2817316368E+00$&$ 0.4731709640E+00$&$\pm 0.4472135955E+01$&$3$\\

$-5$&$ 0.2365739000E+00$&$ 0.3666868582E+00$&$\pm 0.4690415760E+01$&$3$\\
\end{tabular}
\end{ruledtabular}
\end{table}

\section{Conclusion}\label{con}
In this paper we have proposed electron confinement in graphene using smooth magnetic fields which are finite everywhere. Interestingly, when the parameters of the model are subjected to certain constraints the magnetic fields lead to systems which are conditionally exactly solvable. It has also been shown that when these constraints are relaxed it is still possible to determine part of the spectrum analytically, especially the zero energy states. Depending on the orientation of the magnetic field, these states can be found for some values of the magnetic quantum number. For non zero energy states in some cases the algebraic part of the calculations become quite cumbersome and we have obtained the eigenvalues numerically in such cases. It may mentioned that we have examined two values of $N$ in \eqref{eqn:consider-B} but it is possible to consider higher values of $N$. We believe it would be interesting to investigate whether solutions in closed form can be found for a general value of $N$ and if so, what the constraints on the parameters may be. Finally we would like to mention that although we have considered massless electrons it is possible to carry out the entire analysis when a mass term of the form $\Delta\sigma_z$ is present in the Hamiltonian (\ref{eqn:Hamilton}). In such a case the model can be used to study other Dirac materials like silicene \cite{tahir}.

\begin{acknowledgements}
One of the authors (PR) wishes to thank AMOG, Ton Duc Thang University, Ho Chi Minh City for supporting a visit during which this work was carried out.

This research is funded by the Vietnam National Foundation for Science and Technology Development (NAFOSTED) under Grant No. 103.01-2014.44.
\end{acknowledgements}

\end{document}